# Boron−Doping Effects on Local Structures of Semiconducting Ultrananocrystalline Diamond/Hydrogenated Amorphous Carbon Composite Thin Films Fabricated via Coaxial Arc Plasma: an X-ray Absorption Spectroscopic Study

Naofumi Nishikawa[1,2,*,†]

[1] *Department of Applied Science for Electronics & Materials, Kyushu University, 6-1 Kasuga, Fukuoka 816-8580, Japan*
[2] *School of Materials Science, Japan Advanced Institute of Science and Technology (JAIST), 1-1 Asahidai, Nomi, Ishikawa 923-1292, Japan*

\* Corresponding author email: *naofumi_nishikawa@kyudai.jp*

[†] *Present: Independent Researcher*

Ultrananocrystalline diamond/hydrogenated amorphous carbon composite thin films synthesized via coaxial arc plasma possess a marked structural feature of diamond grains embedded in an amorphous carbon and a hydrogenated amorphous carbon matrix which are the largest constituents of the films. Since the amorphous nature yields much larger light absorption coefficients as well as a generation source of photo-induced carriers with UV rays, these films can be potential candidates for deep-UV photodetector applications. From some previous studies p-type conduction of the films has been realized by doping boron in experimental conditions. In addition, their optical and electrical characteristics were investigated previously. However, the bonding structures which largely affect the physical properties of the devices haven't been investigated. In this work, near-edge X-ray absorption fine structure spectroscopy characterizations are carried out. The result reveals that a bonding state $\sigma^*$ C–B of diamond surfaces is formed preferentially and structural distortion is caused at an early stage of boron-doping. Further doping into the films lessens the amount of unsaturated bonds such as $\pi^*$ C≡C, which may be a cause of the device performance degradations. Our work suggests a fundamental case model of boron-doping effects on a local structure of the film.

## I. INTRODUCTION

Diamond thin films, which can be synthesized artificially by various vapor deposition such as hot-filament chemical vapor deposition (HFCVD) [1,2], microwave plasma CVD (MWCVD) [3] and radio-frequency plasma CVD (RFCVD) [4,5], have been studied the past few decades as candidate materials applicable to mechanical and biomedical coating agents, electrodes of fuel cells, semiconductor devices, etc. Since diamond possesses a unique lattice structure with strong covalent bonds among carbon atoms, some intriguing physical properties are original to it. Those are for instance extreme hardness, chemical inertness, excellent thermal conductivity [6,7], dielectric properties of wide-bandgap (5.47 eV) [8]. A lot of applications utilizing these characteristics could be realized by some proper engineering and the methods and processes are developed vigorously even now. Among them, the application of the thin films to semiconductor devices has been longstanding until today. In order to produce semiconducting properties of wide-bandgap materials such as diamonds, impurity atoms that belong to III or V groups are necessary to be doped into the diamond thin films to create acceptor or donor levels between the bandgap. While nitrogen [9], phosphorous [10,11], sulfur [12], and boron [13,14] atoms are cited as the dopant candidates, boron is the most promising dopant that can create p-type conduction in the films since the activation energy is relatively low (0.37 eV) [15,16] so that it can be comparatively easily incorporated into diamond lattice and replaced with carbon atoms.

These potential diamond thin films can be classified by the crystallinity into single crystal and polycrystalline diamonds. While polycrystalline diamonds can be synthesized artificially using hydrogen, methane and/or argon gases via chemical vapor deposition (CVD) technique [17], amorphous fractions and grain boundaries (GBs) between them are inevitably contained in the materials to a greater or lesser extent. This metastable portion is so-called amorphous carbon (a-C) [18] and it often exists as hydrogenated amorphous carbon (a-C:H) in the matter due to the fabrication process. Hence, such polycrystalline diamonds are composites of diamonds and a-Cs or a-C:Hs.

Polycrystalline diamonds are commonly composed of the micro- or nano-structures called microcrystalline diamonds (MCDs) and nanocrystalline diamonds (NCDs), respectively, and the later constituting of several-nanometer diamond grains are sometimes called ultrananocrystalline diamonds (UNCDs) [19], of which composites with an a-C and a a-C:H are written as "UNCD/a-C" and "UNCD/a-C:H", respectively [20].

As stated above, UNCD/a-C and UNCD/a-C:H thin films are deposited on various sorts of substrates via CVDs in general. Meanwhile, these films can also be synthesized via physical vapor deposition (PVD) methods such as pulsed laser deposition (PLD) [21,22] and coaxial arc plasma deposition (CAPD). CAPD process, by which the films are deposited with atomized, ionized, and dimerized carbon particles ejected from a graphite target equipped with a coaxial gun induced with a pulsed voltage [23], is able to form the films without hydrogen gas [24] and diamond seeding pretreatment [25] at comparatively low temperature (550 °C). Therefore, CAPD method can be an environmentally-friendly, low-cost technique. This process also brings about a unique morphology; UNCD grains embedded in an a-C and/or a a-C:H matrices with numerous GBs between them [23]. This distinctive morphology of the films would yield a quite large light absorption coefficient of over $10^5$ cm$^{-1}$ ranging from a photon energy of 3 to 6 eV, two orders of magnitude larger than that of the films by CVDs [23]. This feature is advantageous in photoelectric conversions of UV rays and the UNCD/a-C:H films fabricated by CAPD are great potential candidates for UV-sensing applications (see, for example, Refs. [26] and [27] as for the first experimental demonstrations of functionalities as photo-diodes consisting of boron- and nitrogen-doped UNCD/a-C:H thin films prepared via CAPD, respectively).

A lot of experimental research on the films prepared by CAPD toward the applications to photovoltaics such as UV sensors have been accumulated until today, motivated by some theoretical and experimental studies. Common intrinsic UNCD/a-C:H films synthesized via CAPD possess optical properties such as an indirect optical bandgap of 1.7 eV stemming from the amorphous regions [23], and a direct gap of 2.9 eV owing to GBs specific to UNCD/a-C:H films [23]. The GBs play a crucial part in carrier transport pathway [28], which presumably enables the above applications. Boron-doping into the UNCD/a-C:H films by CAPD has been practically realized by performing deposition using boron-blended graphite targets [26], and the electrical characteristics of boron-doped UNCD/a-C:H films have also been surveyed through p-n heterojunction diodes comprising the boron-doped UNCD/a-C:H films and n-Si substrates [26,28,29]. However, it has not been studied how boron-doping has an influence on the chemical bonding states of the films which possess their complicated structures.

In this work, we report on the boron-doping effects by use of near-edge X-ray absorption fine structure (NEXAFS) spectroscopy, which is functional in quantitative analysis of local structures, studying the boron- and carbon-related bonds since C $K$-edge obtained by this technique is able to decern any bonding states of carbon. Recently, we have investigated the hydrogenation effect similarly by use of NEXAFS spectroscopy as well as X-ray photoelectron spectroscopy (XPS) and Fourier Transform Infrared Spectroscopy (FT-IR), and demonstrated an effectiveness of hydrogenation induced with a technological development of the film fabrication process (see Ref. [30] for the detail) with its detail mechanism by studying comprehensively the chemical bonding state changes accompanied by the amounts of hydrogen content in the films [31]. Results obtained in this work reveal that the films would own double-step transitions of doping toward the bonding states dependent on their amounts and a certain bonding state related to boron surely serves to create electrical levels between the optical gaps closer to the top of valence bands, which leads to a substantial improvement of the electrical conductivities, by correlating with the electrical characteristics investigated previously. Furthermore, the results demonstrate a distinctive hybridization with doped-boron atoms markedly different from those in cases of conventional CVD single and polycrystalline diamonds that the boron dopants predominantly replace substitutionally with carbon atoms that consist of the diamond lattices. The in-depth discussion that would be informative for future works is given and the brief perspectives are also provided finally in Section 4.

## II. EXPERIMENTAL PROCEDURE

We prepared five discrete samples with different boron contents to investigate boron-doping effects on UNCD/a-C:H films in this work. All the samples were fabricated via CAPD based on a conventional manner reported up to date [23,32]. The films were deposited on insulating Si substrates with resistivity of more than 10 kΩ·cm under hydrogen pressure of 53.3 Pa at a substrate temperature of 550 °C after an inside of a chamber was evacuated down up to a base pressure of $10^{-5}$ Pa order by using a turbo molecular pump. Arc plasma discharge was performed with a coaxial arc plasma gun equipped with graphite targets containing 0, 0.1, 1.0, 5.0, 10 at. % boron to form undoped and boron-doped UNCD/a-C:H films. The other important parameters are found in an experimental section of a recent paper [29]. NEXAFS experiments were conducted to obtain B $K$- and C $K$-edges of the spectra in the films at beamline 12 of Kyushu Synchrotron Research Center. All experiments were performed at room temperature in vacuum. The measurement durations were approximately forty minutes in each experiment and the data were recorded at the energy resolutions of 0.05 eV in a photon energy.

## III. RESULTS AND DISCUSSION

First, boron $K$ (B $K$)-edge spectra presented in Figure 1(a) will be studied. These spectra can be divided at a photon energy of ~195 eV into transitions from specific chemical bonds (B–X) to $\pi^*$ and $\sigma^*$ states. The latter, extremely broad plural bands, grow with an increase of boron content in the films. According to some previous research on boron compounds such as boron nitrides (BNs) and boron carbides (BCs), they would be attributed to superpositions of wavefunctions excited from various B 1$s$ orbitals. They are extremely conjugated and it is difficult to carry out a precise investigation on them quantitatively, hence we'll roughly give a discussion on this part. Especially a swell of the spectrum in the region of more than ~200 eV, which can be confirmed in the films with a higher boron content, would be attributable mainly to a feature of an oxidized-boron system as previously reported in the cases of $B_2O_3$ [33,34], BN [34], BN nanotubes (BNNTs) [34] and $B_xC_{1-x}$ [33]. This contribution might be due to the oxide impurities and the similar tendency was also observed in survey-scan and oxygen 1*s* (O 1*s*) XPS spectra (see Ref. [29] and Figure S1 in Supporting Information). The other part, i.e. in a photon energy ranging from ~195 to ~200 eV also originates from excitations of the B 1*s* core-level electron to $\sigma^*$ states. This contribution changes the position of the photon energy by a maximum of up to 4.9 eV based on the material's morphologies as it has been reported in c-BNs and h-BNs [35], which implies that wavefunctions originating from several $\sigma^*$ transitions from different chemical bonds are superpositioned with each other since the UNCD/a-C:H films are composed of a variety of bonding forms especially in an amorphous fraction and GBs, assuming that this description might be applied to our materials.

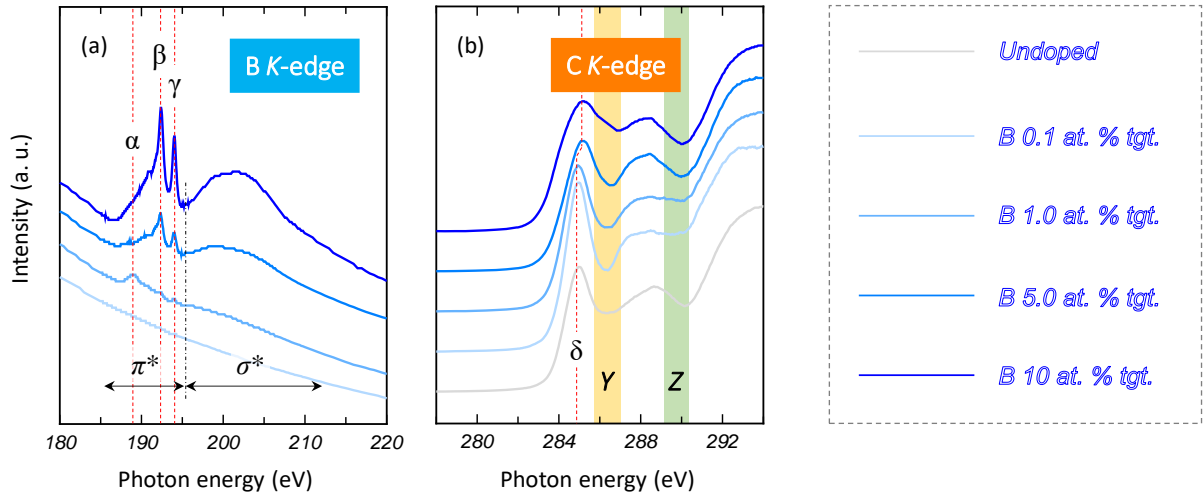

**Figure 1.** (a) NEXAFS spectra in B *K*-edge of UNCD/a-C:H films prepared by use of graphite targets containing 0.1, 1.0, 5.0, 10 at. % boron, and (b) spectra in C *K*-edge of the identical films as well as undoped films synthesized with a pure graphite target. The pre- and post-edges of C *K*-edge are normalized in a proper fashion (see and read the text in Section 3).

Secondary, the area in a photon energy of less than ~195 eV is considered. An absorption spectrum of the film prepared using a target with a boron content of 1.0 at. % exhibits a weak jump at 188.5 eV (labeled α). This would stem from metallic boron (B–B) as demonstrated in a couple of research on h-BNs [36,37], and the left shoulder of the peaks labeled β in the spectra of the films prepared by use of the targets incorporated with 5.0 and 10 at. % boron would also result from this contribution. As a boron content of the film increases, sharp spikes at 192 eV (labeled β) and 194 eV (labeled γ) are becoming noticeable. The former is a typical peak indicating B 1*s* core-level electron to $\pi^*$ states, which is often observed in various sorts of boronide, BN, and BC materials and is specific to them. Especially a literature reporting on B₄C films [38] is suitable for an explanation of this peak corresponding to B–C bonds since this material owns the most similar structure to our films as far as to be known at present. The latter (labeled γ) is weaker as compared to that of the former (labeled β), and there are two considerable possibilities according to some previous studies on BNs and BCs. One is quantum confinement effects that will appear in a scale of less than 20 nm of particle size. As mentioned above, UNCD/a-C:Hs are complex nano-scale materials and are small enough to exhibit such effects, which would be applicable to our films. This hypothesis is supported by a paper reporting on the same observation in BNs [39]. However, it is reported that the effects occur in diamonds whose grain diameters are less than 2 nm [40]. Therefore, it is less possible that quantum confinement effect contributes to a peak labeled γ. Another is an attribution to oxygen impurities as confirmed in some BNs [34,41] and boron powder [42] that might result from surface oxidation. Given that UNCD/a-C:H films synthesized by CAPD generally possess less crystallinity compared with those by CVDs and most of them is composed of disordered amorphous structures and GBs, quantum size effects induced with UNCD grains would be limited. Hence, most of these peaks labeled γ are presumed to be ascribed to the oxide impurities of the films and the quantum confinement effects will be the slightest contributions. To summarize the studies of B *K*-edge, boron-doping will bring about the following quantitative variations; (i) at an early stage of doping, metallic boron is formed in the films and (ii) as the amount of dopant increases, boron atoms are certainly incorporated to form B–C bonds in association with an involvement of oxygen impurities, resulting in a formation of boron-oxides.

To investigate how the carbon hybridization in the films varies quantitatively by doping boron, NEXAFS C *K*-edge spectra were examined at the same beamline and analyzed in detail. Each extracted data was processed based on some previous studies on UNCD/a-C:H films accumulated up to the present (see, for example, Refs. [43] and [44] regarding the first investigation of the spectra and well-fitted data processing, respectively). The backgrounds were subtracted and pre-edge in the region of less than 283 eV and post-edge more than 314 eV of the spectra were normalized to reveal original absorption. Afterwards, arctangent function steps are applied as an approximation of the largest absorption located at ~291 eV that denotes the ionization potential peculiar to diamonds. As it has been already studied on our recent paper, the NEXAFS C *K*-edge can be sorted out on the basis of the derivations; specifically, the areas larger than ~291 eV come from UNCD grains, an a-C and a a-C:H matrices [31]. It should be noted here that the threshold of this absorption differs depending on the morphologies of any carbon-related materials (see, for instance, Ref. [45] regarding the comparative work). Since this fraction (over ~291 eV) seems not to give rise to any new findings, we will discuss the topic excluding this area.

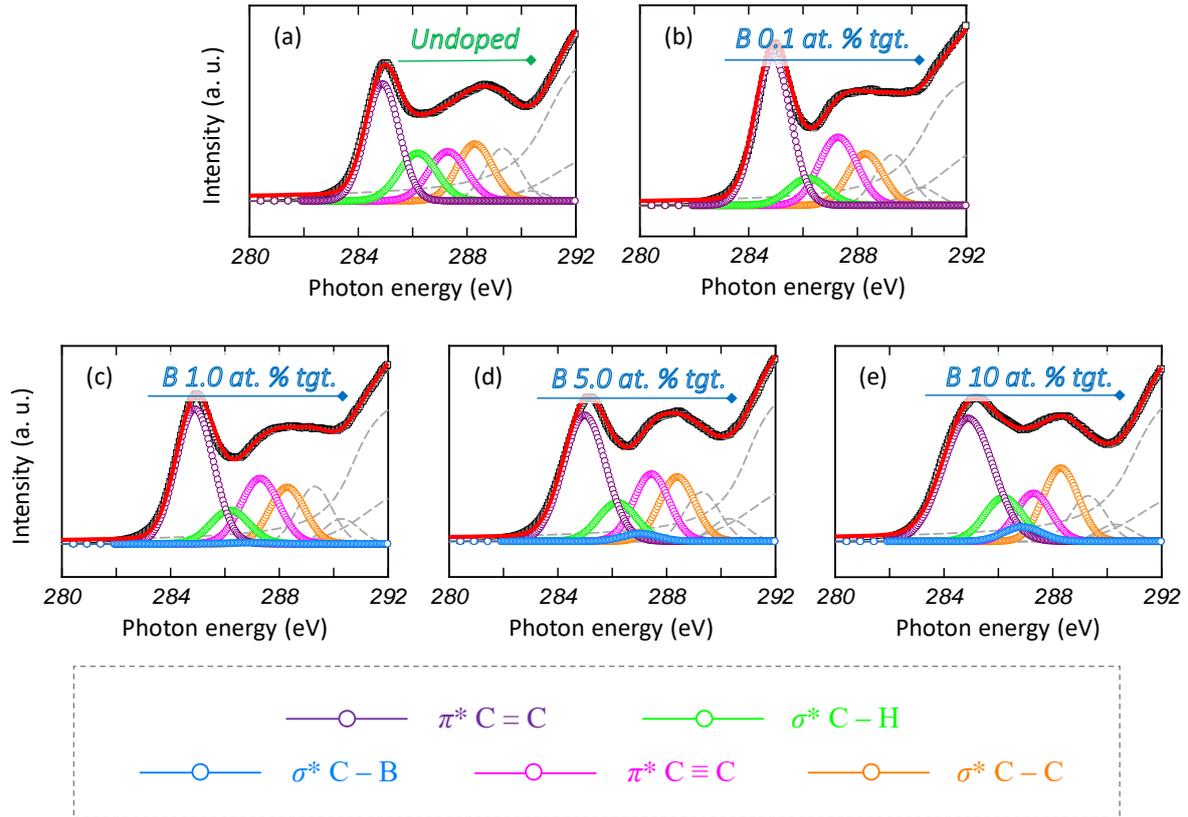

**Figure 2.** Magnified NEXAFS C *K*-edge spectra of (a) undoped and (b)–(e) boron-doped UNCD/a-C:H films fabricated using graphite targets incorporated with (a) 0, (b) 0.1, (c) 1.0, (d) 5.0, (e) 10 at. % boron (indicated with black squared dots). Each spectrum is deconvoluted with gaussian and arctangent functions representing each carbon hybridization and large X-ray absorption owing to UNCD crystallites, respectively, and the sum totals are exhibited with red curves. As for the former, important contributions in this work are colored with purple ($\pi^*$ C=C), light green ($\sigma^*$ C–H), blue ($\sigma^*$ C–B), magenta ($\pi^*$ C≡C) and orange ($\sigma^*$ C–H), and the others are expressed with dashed curves. All the details of a photon energy except for the case of $\pi^*$ C=O and $\pi^*$ COOH are referred to our latest paper [31]. These spectra were obtained by Total Electron Yield (TEY) technique, different from the referred work, and processed in a proper manner; therefore, note that both figures cannot be compared simply with each other.

These spectra processed in the above stated fashion are presented in Figure 1(b). A distinct peak positioned at the lowest photon energy, which comes from $\pi^*$ C=C resonance and is labeled δ, denotes energy shift toward higher energy by doping boron at a transition point between the case using a B 1.0 at. % target and a B 5.0 at. % one. This observation might be an implication of a formation of carbon oxides deriving from oxidation process accompanied by boron-doping. Since a formation of B–C bonds from C–C bonding system hardly affects the NEXAFS spectra as it has been reported any number of times previously in some sorts of BCs, it's the most reasonable interpretation.

One of the most noticeable appearance changes are illustrated by the areas of *Y* and *Z* in Figure 1(b). To study the details of these areas, the processed spectra were decomposed to gaussian functions that represent X-ray absorption in each carbon hybridization by use of analytical fitting program Athena [46] based on our recent work [31] and they are shown in Figure 2. Specifically, backgrounds of the obtained raw data were subtracted at the first setout, and the pre- and post-edges of processed data were normalized. Deconvolution was carried out, on the basis of some previous paper reporting the successful processing, with a careful attention to fitting parameters such as peak position and full width at half maximum (FWHM). According to a previous research on boron-doped UNCD/a-C:H films synthesized via PLD technique investigated by NEXAFS spectroscopy, $\sigma^*$ C–B resonance in the spectra comes out at a photon energy of 286.9 eV [47], while the similar spikes were observed in CAPD UNCD/a-C:H films by boron-doping [48]. Though the feature is not to be confirmed in Figure 2(a) and 2(b), namely in the undoped and 0.1 at. % boron-doped films, it stands out gradually from Figure 2(c) to 2(e), i.e., with an increase of boron content in the films (see also the part set with a blue net of Figure 3), which means a successful incorporation of boron to hybridize with carbon atoms into formation of C–B bonds in the films. Flattening transition of the spectra by doping boron in the area of $Y$ in Figure 1(b) would be caused mainly by this chemical bonding formation. Secondly, we shall place attention to the deconvoluted spectra colored with light green in Figure 2. This contribution coming from $\sigma^*$ C–H and centered at 286.2 eV suddenly decreases when the slightest amount of boron is doped as illustrated by a light green net of Figure 3. Once the boron is incorporated into the films, this hybridization gradually increases accompanied by an increase of the dopant, and the similar tendency can be observed in $\sigma^*$ C–C though it exhibits less conspicuous compared with that of $\sigma^*$ C–H. Since both of them are saturated bonds, it is estimated that incorporated boron atoms dismantled these stable structures with its extremely small amount, and such distorting effects are being mitigated while the boron-doping is assisted. Concerning the excitation to $\pi^*$ C≡C states, a totally opposite inclination can be confirmed in Figure 3 (see the columns with a magenta net). It can be interpreted that $\pi^*$ C≡C, the most unstable state, is generated additionally by boron-doping at the initial stage owing to the above-mentioned structural distortion, and in the second doping phase with a boron concentration of the target more than 1.0 at. % this triple-bonded system is stabilized by addition reaction to form more stable states such as $\sigma^*$ C–H, $\sigma^*$ C–C, and $\pi^*$ C=C as well as $\sigma^*$ C–B as it can be confirmed in Figure 3. The correlation with oxygen impurities is unclear at present and should be investigated by conducting measurements on NEXAFS O $K$-edge. The amount of $\pi^*$ C=C in the films increases approximately accompanied by additional boron doping, whereas it represents an irregular picture in the case of the films deposited with a boron content of 0.1 at. % in a target. This is presumed that the light-doping of boron causes the structural-distortion effect conspicuously, resulting in more $\pi^*$ C=C states detected, and the effect is promoted roughly proportional to the amount of dopant thereafter.

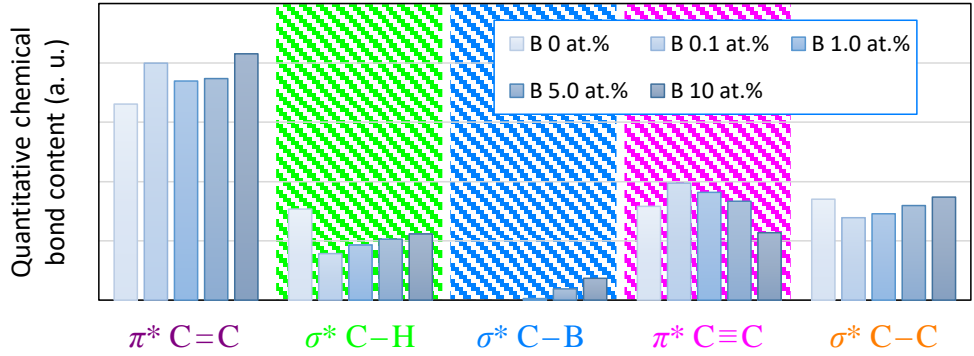

**Figure 3.** Relative amount of each carbon hybridization in boron-doped and undoped UNCD/a-C:H films synthesized by use of graphite targets with a boron content of 0, 0.1, 1.0, 5.0, and 10 at. % obtained from the deconvolutes of NEXAFS C $K$-edge presented in Figure 2.

The decomposed spectrum deriving from carbon-oxygen bonds is split into two gaussian functions positioned at 289.3 and 290.3 eV which denote $\pi^*$ C=O and $\pi^*$ COOH, respectively, from that in reference to our latest work [31]. The area around here is highly convoluted with several absorptions such as an ionization potential of UNCD crystallites and a core electron excitation to $\sigma^*$ C=C states as well as that denotes a transition of $\pi^*$ C=O and $\pi^*$ COOH. Hence, quantification of these $\sigma^*$ states is susceptible to the fitting parameters and lacks its accuracy, and it's not to be discussed anymore in this work since it never affects the discussion on the other parts considering comprehensively the FWHMs of each deconvolutes equivalent to 1.53 eV and the positions enough apart from each other. This portion in the deconvolution process of NEXAFS C $K$-edge is under development at present and requires extra studies to comprehend the details more precisely. For instance, Shpilman et al. reported the existence of $\pi^*$ C=O at ~286.5 eV in CVD diamonds [49], though it is not considered here as there have been a lot of works on PVD UNCD/a-C:H thin films including CAPD ones, published in actual and yielding many intriguing results by quite well fitting process almost the same with that I used in this work.

These chemical bonding state transitions unveiled by NEXAFS B $K$- and C $K$-edges summarize comprehensive mechanisms of boron-doping effects on the local structures of UNCD/a-C:H films deposited by CAPD technique. At an early stage of doping with its amount of less than 1.0 at. %, incorporated boron atoms form metallic boron without any hybridization with carbon atoms. This state of boron would be localized at GBs predominantly as reference to a report on boron-doped UNCD/a-C:H films [50] that were fabricated the most similarly to ours as far as comparable with our results. The doped-boron doesn't replace substitutionally with carbon atoms comprising diamond lattice, much less exist as metallic in the diamond grains, and it also be less possible that it is present inside an a-C and/or a-C:H matrices without bonding with carbon atoms, considering their kinetics. This localization gives rise to structural strains and unstable states such as $\pi^*$ C=C and $\pi^*$ C≡C that would be transformed from more stably saturated states especially from $\sigma^*$ C–H of which hydrogen plays a part in termination of dangling bonds at GBs. Further doping makes the dopants combined with carbon atoms to form $\sigma^*$ C–B. It is considered that the doped-boron would replace mainly hydrogen atoms substitutionally and append partially to dangling bonds, both of which are situated at the UNCD surfaces exposed to GBs, similarly to the previous investigations on the structures of PLD boron-doped UNCD/a-C:H films [47]. Simultaneously oxygen impurities are involved with boron to generate boron oxides presumably localized at the film surface, based on a result of an ion bombardment etching treatment in an XPS survey [51]. Incorporated-boron will permeate the whole films uniformly [26] to form the above-mentioned bonding state, resulting in the enlargement of UNCD grains [48,52] of which size is presumed to be tens of nanometers, so-called NCDs. Concretely, the doped-boron doesn't incorporate into the lattice of UNCD crystallites and replace with the constituent carbon atoms, but substitute preferentially with hydrogen atoms terminating the UNCD surfaces at GBs. This doping mechanism is totally different from phenomena confirmed in a lot of conventional CVD diamonds introduced in Section 1, and is rather similar to those in PLD UNCD/a-C:H thin films. The grown diamond nanoparticle increases its concentration in the films, which generally decreases the number of, but might thicken the GBs as reported in the case of nitrogen-doped CVD UNCD thin films [53]. These denser diamonds would serve to generate more photo-induced carrier under UV irradiations, therefore the diodes comprising the films will be more advantageous in applications to UV photodetectors.

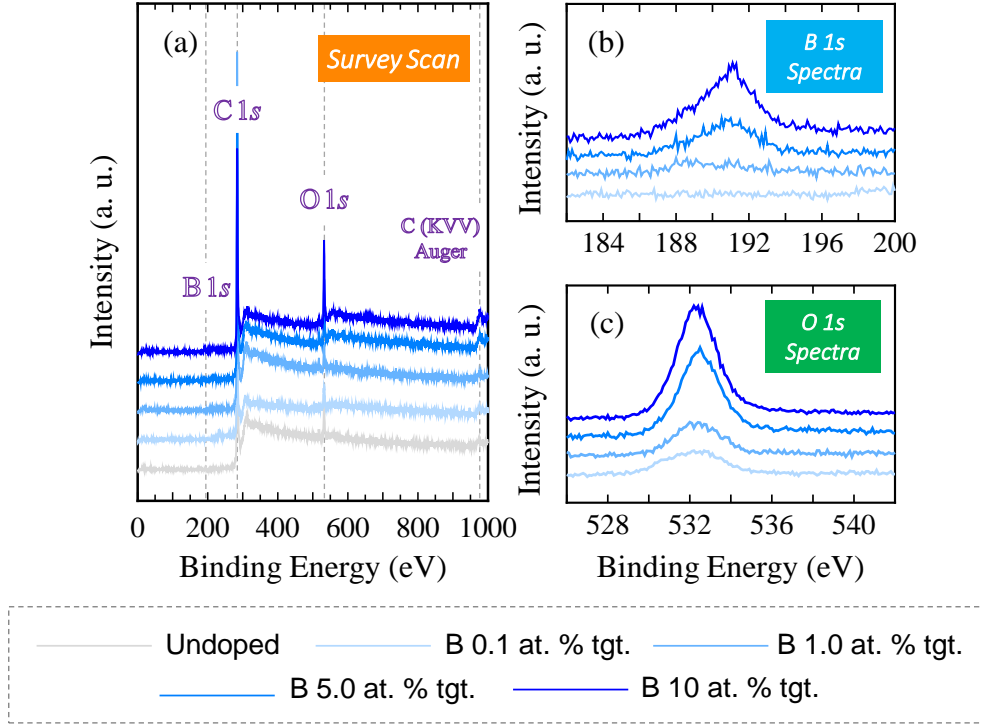

**Figure S1.** X-ray photoelectron spectroscopy of (a) survey-scanned, (b) boron 1$s$ (B 1$s$), and (c) oxygen 1$s$ (O 1$s$) spectra. All measurements were conducted at beamline 12 of Kyushu Synchrotron Light Research Center by use of soft X-rays. In parallel, Mg K$\alpha$ line with a photon energy of 1253.6 eV was utilized for the survey scan measurements. Survey scan spectra are typical figures of UNCD/a-C:H films and a couple of jumps centered at around 285 eV, 532 eV, and 980 eV that come from carbon 1$s$ (C 1$s$), B 1$s$ orbitals and carbon KVV auger transition in order are observed. The details of each constituent are studied in a recent work [31]. B 1$s$ spectra show an advent of a peak, which means certainly boron is incorporated into the films. Growing O 1$s$ spectra accompanied by boron-doping indicate the oxygen is involved at the film surface with an increase of incorporated-boron.

## IV. CONCLUSIONS

By use of NEXAFS spectroscopic technique, boron-doping effects on UNCD/a-C:H films deposited by CAPD have been studied to clarify the chemical bonding transitions of their local structures. The results indicate metallic boron is formed sparsely at GBs to exist at interfaces between UNCD crystallites and an a-C and/or a a-C:H matrices at an early stage of doping, and $\sigma^*$ C–B starts to be generated on the UNCD grain surfaces substitutionally with hydrogen atoms at the surfaces on a priority basis and dangling bonds subsidiarily from a certain point by the additional doping. This chemical bonding states should be an origin of a drastic enhancement of electrical conductivities of the films and play a part in acceptor-like levels localized above the valence band, which would be helpful in inducing hopping conductions [28]. Our results and in-depth investigations that suggest an empirical model of boron-doping on the films would be a fundamental platform of future works of the films and open the way for relevant research toward various applications requiring impurity doping, which are represented by semiconducting devices such as photovoltaics.

### Note and Acknowledgments

### Conflict of Interest

The author declares no conflict of interest.

### Data Availability Statements

The data that support the findings of this study are available upon reasonable request from the authors.

### Appendix. Additional Discussion

We provide here additional discussion on electrical characteristics that the UNCD/a-C:H films possess revealed in some previous research and correlate them with the results obtained this time to draw brief perspectives on a possible application toward UV photodiodes. Electrical properties and transport mechanisms of the films in each doped level can also be explained simultaneously by comparing the above discussion with some previous research. Hereupon, we review the outcomes regarding the films accumulated up to now before initiating correlation. As introduced in Section 1, heterojunction diodes comprising boron-doped UNCD/a-C:H films and n-type Si substrates were successfully fabricated with CAPD technique and the electrical characteristics as well as the transport mechanism have also been investigated. It should be noted that CAPD UNCD/a-C:H thin films occupy their constituents mostly of a-C and a-C:H matrices accompanied by a great number of GBs differently from those fabricated with various conventional CVD techniques which own denser polycrystalline diamonds with less other components such as amorphous and GBs. This compositional difference may have generated their distinguished optical and electrical properties represented by the three different optical gaps deriving from a-C and/or a-C:H matrices, GBs, and diamond particulates. As for the electrical transport mechanism, it has been explained that firstly electron-hole pairs are generated at UNCD grains with UV irradiation. After that the created electron and hole transfer through mainly GBs in addition to their amorphous fractions supported by incorporated nitrogen and boron impurities in n- and p-type CAPD UNCD/a-C:H films, respectively. Recently, Katamune *et al.* reported the electrical characteristics of boron-doped UNCD/a-C:H films by use of hard X-ray photoelectron emission microscopy (HAXPES). From an accompanied investigation, it is known that the activation energy of the films rises with a drop of boron content and that more doped films with the amount of dopant more than 2.0 at. % exhibited clear tendencies in a hopping model [54], which was revealed from the temperature-dependence of electrical conductivity measurements. Besides, it reports boron-doping into the films makes the Fermi level move toward to the valence band and forms acceptor-like localized levels, which induces the three-dimensional variable range hopping (VRH) conduction, with lowering the total electron density of state. These details can be referred to Ref. [29].

In lightly-doped films with its amount of presumed 0.5 at. %, photo-induced carriers generated at UNCD crystallites with UV illuminations would pass mainly through GBs with an assistance of metallic boron via three-dimensional VRH and secondary through an a-C and a a-C:H matrices as discussed in our recent work [31]. Though undoped CAPD UNCD/a-C:H films possess less conductivities equivalent to $2 \times 10^{-7}$ $\Omega^{-1}$cm$^{-1}$ at room temperature [29], p-n heterojunction diodes consisting of the lightly-doped films and n-type Si substrates surely exhibited the rectifying actions [30] and work sufficiently to be p-type semiconductors. The films doped with an amount of boron enough to form $\sigma^*$ C–B should increase their electrical conductivities up to $2 \times 10^{-1}$ $\Omega^{-1}$cm$^{-1}$ at room temperature owing to the created localized states in the optical gap mentioned above, which leads to an improvement of carrier concentration. Diodes with the films and the n-Si substrates also follow the VRH and photo-induced carriers may transport similarly to those with a small amount of boron incorporated. In any cases the carrier paths and the transport mechanisms are the same much or lesser, and the detail descriptions are provided in Ref. [31]. We should note here that oxygen impurities involved accompanied by this process hardly have an influence on the electrical conductivities referred to Ref. [29].

Thus, we have conducted close correlation between chemical bonding states and electrical properties of boron-doped UNCD/a-C:H thin films prepared via CAPD in each doping level in this section. The result to be emphasized the most is a bonding state $\sigma^*$ C–B would be an origin of enhancing electrical conductivities of the films. However slight the quantitative variation is, it surely serves to improve device performances in the p-n heterojunction diodes.